%% file: Main.tex
\documentclass{article}

\input{preamble}

\usepackage{natbib}

\usepackage{authblk}

\usepackage{hyperref}

\usepackage{fullpage}

\title{libRoadRunner: A High Performance SBML Simulation and Analysis Library}

\author[1]{Endre T. Somogyi}
\author[2]{Jean-Marie Bouteiller}
\author[1]{James A. Glazier}
\author[3]{Matthias K\"{o}nig}
\author[4]{Kyle Medley}
\author[1]{Maciej H. Swat}
\author[4]{Herbert M. Sauro}

\affil[1]{Biocomplexity Institute and Department of Physics, Indiana University, Bloomington, IN
  47405, USA}
\affil[2]{Biomedical Engineering Department, University of Southern California, Los Angeles, CA 90089, USA}
\affil[3]{Department of Computational Systems Biochemistry, University Medicine Charit\'{e}  Berlin, 10117 Berlin, Germany}
\affil[4]{Department of Bioengineering, University of Washington, Seattle, WA 98195, USA}

\begin{document}

\maketitle

\begin{abstract}

\section{Motivation:} 
This paper presents libRoadRunner, an extensible, high-performance, cross-platform, open-source software library for the simulation and analysis of models expressed using Systems Biology Markup Language (\textit{SBML}). SBML is the most widely used standard for representing dynamic networks, especially biochemical networks. libRoadRunner supports solution of both large models and multiple replicas of a single model on desktop, mobile and cluster computers. 
  
\section{Results:}
libRoadRunner is a self-contained library, able to run both as a component inside other tools via its C++ and C bindings andnteractively through its Python interface. The Python Application Programming Interface (\textit{API}) is similar to the APIs of  Matlab and SciPy, making it fast and easy to learn, even for new users. libRoadRunner uses a custom Just-In-Time (\textit{JIT}) compiler built on the widely-used LLVM JIT compiler framework to compile SBML-specified models directly into very fast native machine code for a variety of processors, making it appropriate for solving very large models or multiple replicas of smaller models. libRoadRunner is flexible, supporting the bulk of the SBML specification (except for delay and nonlinear algebraic equations) and several of its extensions. It offers multiple deterministic and stochastic integrators, as well as tools for steady-state, stability analyses and flux balance analysis.

\section{Availability and Implementation:}
We regularly update libRoadRunner binary distributions for Mac OS X, Linux and Windows and license them under Apache License Version 2.0. \href{http://www.libroadrunner.org}{http://www.libroadrunner.org} provides online documentation, full build instructions, binaries and a git source repository.  \section{Contacts:} \href{hsauro@u.washington.edu}{hsauro@u.washington.edu}, \href{somogyie@indiana.edu}{somogyie@indiana.edu}

\end{abstract}

\section{Introduction}

Dynamic network models~\citep{SauroBookPathwayModeling} of metabolic, gene regulatory, protein signaling and electrophysiological models require the specification of components, interactions, compartments and kinetic parameters. The Systems Biology Markup Language (\textit{SBML})~\citep{Hucka:2003fs} has become the \textit{de facto} standard for declarative specification of these types of model (see SBML.org and refs.).

Popular tools for the development, simulation and analysis of models specified in SBML include COPASI~\citep{Hoops:2006ui}, Systems Biology Workbench (\textit{SBW})~\citep{Bergmann:2006ut}, The Systems Biology Simulation Core Algorithm (\textit{TSBSC})~\citep{Keller:2013cn}, Jarnac~\citep{sauro2000jarnac}, libSBMLSim~\citep{Takizawa:2013tj}, SOSLib~\citep{url:soslib}, iBioSim~\citep{iBioSim:2009}, PySCeS~\citep{Pysces:2005}, and VirtualCell~\citep{Moraru:2008iv}. Some of these applications are stand-alone packages designed for interactive use, with limited reusability as components in other applications. Very few are flexible component libraries. Currently, none are fast enough to support emerging applications that require large-scale simulation of network dynamics. For example, multicell virtual-tissue simulations~\citep{Hester:2011tz} often require simultaneous simulation of tens of thousands of replicas of models residing in their cell objects and interacting between cells. In addition, optimization methods require generation of time-series for tens of thousands of replicas to explore the high-dimensional parameter spaces typical of biochemical networks~\citep{bouteiller:2014}.

We designed libRoadRunner to provide: 1) Efficient time-series generation and analysis of large or multiple SBML-based models; 2) A comprehensive and logical API; 3) Interactive simulations in the style of IPython and MATLAB; and 4) Extensiblity.

Most existing SBML simulation engines use built-in interpreters to parse and execute SMBL model specifications. Interpreted execution is simple and flexible, but much slower than execution of compiled code. Other simulation engines generate compiled executables from SBML by first converting SBML-specified models into a general-purpose-language representation. The engines then call an external compiler to translate the general-purpose-language into an executable shared library to load at run time.  \textit{E.g.}, roadRunner in the SBW suite~\citep{Bergmann:2006ut} converts SBML into C\# (see~\S~1.4 of~\citep{Aho:1986}), then compiles the C\# using a compiler from the .NET distribution. This approach generates relatively fast executables. However, it requires distribution of a separate compiler or a redistributable runtime, reducing portability. 

A more efficient approach to SBML-to-executable compilation uses a specialized \emph{just in time}~\textit{JIT} compiler, to compile SBML into an optimized Intermediate Language (\textit{IL}) representation and the IL code into native executable machine code directly in-memory. Ackermann \textit{et al.}~\citep{Ackermann:2009vq} used JIT compilation to generate CUDA code from SBML and execute it on an Nvidia GPU. libRoadRunner supports execution of a broad range of SBML models on CPUs using a custom-built JIT compiler (based on the LLVM JIT compiler framework ~\citep{Lattner:2004vw}) which translates SBML into highly-optimized executable code for a broad range of processors. LLVM-based compilers are small, so all JIT operations occur in memory, without external file or compiler access, ensuring fast, self-contained simulations and a relatively small distribution package.

\paragraph{Capabilities} 
libRoadRunner supports time-course simulation or deterministic and stochastic models. In addition it supports  steady state, stability analysis and flux balance analysis of SBML-specified models. For flux balance analysis libRoadRunner support the FBC SBML extension~\citep{url:sbmlfbc}. It supports almost the entire SBML L3V1 specification, including composite model composition and the distribution package. Its only non-compliance is its lack of support for delay equations and non-linear algebraic rules.

\paragraph{Portability} 
Because new hardware platforms appear frequently, a modern simulator must be portable. libRoadRunner has no run-time dependencies beyond standard system libraries and it supports any processor LLVM supports. libRoadRunner is  written in C++, so it interfaces easily with other C++-based software. libRoadRunner also provides a C language wrapper for cross-language support, and uses SWIG~\citep{beazley1996swig} to provide a customized native-Python API. The use of SWIG will allow future support for additional native language bindings, such as JavaScript, R, or MATLAB, depending on demand.

\paragraph{Extensibility}
libRoadRunner's modular design is easy to maintain and extend. All top-level components, such as solvers and integrators, interact via well-defined boundaries (\emph{pure virtual interfaces}) to reduce inter-component dependencies and hide their internal details. A new solver only
needs to implement a standard interface to function as part of the library, so adding a solver requires no modification of pre-existing code.

\subsection*{Systems Biology Markup Language as a Declarative Language}
SBML~\citep{Hucka:2003fs} is a \textit{declarative} specification format for network models. Because of its history, SBML terminology derives from biochemistry and includes common biochemical-reaction abstractions like reaction steps, compartments and reaction rate laws, though it can describe any model of form: 
\begin{equation}
  \frac{d}{dt}\mathbf{x}(t) =
      f(\mathbf{x}(t), \mathbf{p}),
  \label{eqn:sbmleqn}
\end{equation}
where $\mathbf{x}$ is the state vector of the model, and $\mathbf{p}$ is a vector of time-independent parameters. 

SBML-specified models can also include \textit{events}, discontinuous state changes, which trigger under specified conditions. lib\-Road\-Runner correctly handles SBML-specified events and extends the SBML specification by allowing an SBML event to call an arbitrary user-defined function.

Declarative specification languages, like SBML, define component objects and their interactions, rather than defining procedural control flow (\textit{i.e.}, the sequence in which computational operations proceed on execution). An SBML specification lists only the network component objects, their interactions and rate relations and events which change these interactions and rates, all of which are intrinsic abstractions in SBML. Thus, an author writing a model specification in SBML can focus on the underlying biology or chemistry of the model rather than on how to implement the model as a simulation. Since SBML does not specify the computational operations to implement a model, the control flow, the solvers to use, or how to store the model's elements, an SBML compiler or interpreter must generate them appropriately from the SBML specification. Thus, compiling an SBML model specification is more complex than compiling a functionally-equivalent model specification in a procedural language. 

SBML model specifications are easier to share than procedural specifications of equivalent models because they are not impl\-ementat\-ion dependent; any of the numerous SBML compliant tools can process any SBML model specification. This portability allows model archiving (\textit{e.g.}, in exchange repositories such as BioModels~\citep{url:biomodels}) and reuse and the relatively simple assembly of multiple SBML-specified submodels into larger models. It also simplifies the scientific validation of SBML-specified models and ensures that SBML-specified models remain usable, even if the specific software tools that generated them fall out of use.

\section{Architecture} \label{sec:arch}
libRoadRunner is a self-contained, easily embedded library with an object-oriented API natively accessible in C, C++ and Python (SWIG allows easy extension to other languages). libRoadRunner's \emph{component-oriented design} specifies a small number of standardized software interfaces (protocols) and how they interact, implemented using standard C++ data types. Component-orientation separates the implementation of a component from its interface, so components are easy to add or replace and component swapping requires no changes to existing code. \textit{E.g.}, we can add new integrators, steady-state solvers or SBML compilers to the libRoadRunner library via the \texttt{Integrator}, \texttt{SteadyStateSolver} and \texttt{ExecutableModel} interfaces respectively. libRoadRunner incl\-udes three implementations of the \texttt{Integrator} interface: two deterministic integrators (one based on the CVODE integrator from the Sundials suite~\citep{hindmarsh2005sundials} and the other a standard fourth-order Runge-Kutta method) and a standard Gillespie SSA stochastic integrator. libRoadRunner implements the \texttt{SteadyStateSolver} interface as a class which uses the NLEQ~\citep{nowak1991family} solver, and we are currently developing additional methods. libRoadRunner implements the \texttt{ExecutableModel} interface as a class which uses our SBML-to-CPU JIT compiler (see \S~\ref{sec:sbml_compile}), and we are developing an SBML-to-GPU JIT compiler as well. libRoadRunner statically links to the third-party libraries LLVM~\citep{Lattner:2004vw}, libSBML~\citep{bornstein2008libsbml}, CVODE, NLEQ2, LAPACK and Poco.

\section{SBML-to-CPU-Executable Compilation} \label{sec:sbml_compile}
LibRoadRunner's SBML JIT compiler compiles SBML models in the form of strings to executable native
machine code, in memory. Compilation follows the canonical compiler phases~\citep{Aho:1986}:
(1) lexical analysis, (2) syntactic analysis, (3) semantic analysis, (4) intermediate code generation, (5) code optimization, and (6) native code generation. Standard generic libraries can perform phases 1, 2, 5 and 6. However, semantic analysis (phase 3) is specific to the source language. 

In phases 1 and 2, the compiler reads the source text, parses it, and extracts and converts the text's syntactic information into an \emph{abstract syntax tree} (\textit{AST}) data structure. Each node in the AST is an \textit{essential construct} such as an operator, symbol, literal or function call. Most SBML simulators use components of the libSBML~\citep{bornstein2008libsbml} library to perform lexical and syntactic analyses of SBML model specifications. 

In phases 3 and 4, the compiler reads the AST and assembles it into a sequence of \emph{intermediate language} (\textit{IL}, a machine-independent assembly language) instructions which form a procedural instantiation of the SBML model specification. 

CPUs cannot execute IL programs directly, so phases 5 and 6 optimize the IL (by removing redundant operations, optimizing memory layout,...) and convert it into executable machine code. libRoadRunner uses components of the LLVM library for phases 5 and 6. An advantage of the LLVM code-generator is that it can produce executables for x86, GPUs, ARMs and other processors. 

After the completion of phases 1-6, the JIT compiler returns the executable code in the form of a list of callable functions to the calling program. 

During phase 3 (semantic analysis), the compiler must map language symbols to memory address locations. The compiler of a procedural language such as C, allocates a memory location to each symbol (\textit{e.g.}, a variable or function declaration), and resolves that symbol to that location whenever the source code references that symbol. Procedural-language compilers map symbols to memory locations using a \emph{symbol table} data structure. SBML has no construct for creating new variables or eliminating variables at run-time, so the compiler can compute the exact memory requirements for all symbols and store the symbols in a contiguous memory block. At run-time, during a time-series computation, the libRoadRunner library connects a JIT-compiled function to an integrator, which, in turn, calls a function which calculates the rate of change of the state vector. Since both the state vector and the rate of change occupy contiguous memory blocks and have the same layout as the SBML model variables, the calls pass only two pointers and require no memory copying or rearrangement.

However, compilation of SBML poses challenges. SBML model specifications may define rules which state that an expression should replace a specified symbol, or a rate rule which specifies a rate of change of the value of a symbol, rather than the symbol value itself. SBML also allows different rules to apply in different contexts, such as special rules which only apply when the model is loaded (initial assignment rules). Mapping symbol names to memory locations is not one-to-one so a symbol table is insufficient to store the mapping.

Some SBML model simulators allocate storage space for both normal and rule-defined symbols and use auxiliary functions to evaluate the rules at run-time as the symbols are read. However, this approach wastes memory storing symbols which resolve to other symbols and complicates execution, as the run-time must keep track of rule dependencies. 

Our solution is to extend the symbol table into a \emph{symbol forest}, a hash table which maps symbol names to ASTs describing all the symbols' rules. The SBML compiler uses the symbol forest much as a procedural-language compiler uses a symbol table, to resolve symbol names to memory locations. However, the symbol forest must apply any rules which relate symbols to determine the memory location for a given symbol. \textit{E.g.}, if the symbol $x$ has the assignment rule $x \to y + 1$, whenever the compiler references symbol $x$, the symbol forest will find the rule, generate a sequence of IL instructions which both implement the right hand side (\textit{RHS}) of the rule and create a temporary variable to store the result of the rule calculation. The symbol forest then stores this sequence of IL instructions and returns the memory location of the instruction sequence to the compiler. Later in compilation, the LLVM code generator translates these IL instructions into an executable, which calculates and returns the value of the symbol at run-time. The symbol forest resolves automatically recursive rules in which the symbols in the RHS of a rule depend on other rules. 

Na\"{i}vely generating IL expansions of the rule definitions inline and creating temporary variables for rule evaluation would generate redundant instructions which would slow both compilation and execution. libRoadRunner's \emph{scoped symbol cache} reduces such redundancy. Many functions in libRoadRunner do not modify SBML-defined parameters and variables during function execution, so any rules depending on these parameters and variables need evaluation only once during a given call to these functions. Even if the rule involves a condition, \textit{e..g.}, $x \to \{ b \; \text{if}\; (a > 1)\; \text{else}\; c)$, if the function does not change the values of $a$, $b$ and $c$, the function need evaluate the rule to obtain the value of $x$ only once per call. The SBML compiler therefore generates code which evaluates the rule whenever the function is called and stores the result in a temporary variable. During a call to the function, the first reference to the symbol evaluates the rule and caches its result, and any subsequent references to that symbol during the function call reference the cached value. Using a scoped symbol cache reduces both memory usage and execution time, typically by a factor of 10 for large models. 

When JIT-compiled functions contain conditional branches which contain rules, the SBML compiler generates redundant IL code, which slows compilation (which scales as the size of the IL code) but has no speed cost at execution. If the compiler examined all possible branches, determined what rules were present, and created temporary variables to contain the results of the rule evaluations, it would reduce the size of the resulting IL code, speeding compilation. However, slower execution would offset the faster compilation, since the executable would evaluate all rules in all branches, not only those which it needed. We may add a compiler directive to allow the user to choose the second option in a future release of libRoadRunner. 

\section{Results}
\subsection*{Performance}
Simulation engines which interpret SBML models~\citep{Romer:1996wj}, are inherently slower, sometimes much slower, than engines which genearte and execute complied code. libRoadrunner uses JIT compilation to generate particularly fast simulations.

Simulation speed depends on the performance of both the state-vector rate calculation and the numeric integrator. Since we cannot separate these calculations in most SBML-model packages, we compared an SBML model JIT-compiled using libRoadRunner with a hard-coded C++ version of the same model. The model implemented 1000 instances of a unimolecular reaction, in which a single substrate goes to a single product (\ce{S -> P}) at a rate of
\begin{equation*}
    \frac{\text{Vm} \displaystyle\left(\frac{S}{\text{Km}_1}\right) \left(1 - \frac{\Gamma}{K_{eq}}\right) \left(\frac{S}{\text{Km}_1} + \frac{P}{\text{Km}_2}\right)^{h-1}}
         {\displaystyle\frac{1 + (M/k)^h}{1 + \sigma*(M/k)^h} + \displaystyle\left(\frac{S}{\text{Km}_1} + \frac{P}{\text{Km}_2}\right)^h}.
\end{equation*}
This equation models a reversible Hill equation~\citep{SauroBookEnzymeKinetics} that was developed by Hofmeyr and Cornish-Bowden~\citep{ReversibleHill:1997}. This test measured the total simulation time for performing between 1000 to 15,000 time steps, and on average, the JIT compiled SBML model performance was 96\% that of a hard-coded C++ specification. All tests were performed on a 64 bit Linux system, and clang was used as the C++ compiler. The timing data implies that we are close to optimal performance. That is, other than using a completely different approach, such as parallelization, it is not possible to improve performance further. 

\subsection*{Python Bindings}
The Python API employs a simple, concise object model, and follows the style and conventions of the widely-used SciPy library which makes it easy to learn. The API also provides high performance, low-overhead access to the libRoadRunner library. The API only communicates using standard Python data types such as lists, dictionaries and NumPy arrays, which simplifies integration with existing applications. The NumPy array type is a data structure which wraps a Python interface around a standard C numeric array. Even large NumPy arrays have low overhead, since they return only pointers to internal arrays owned by the libRoadRunner library, with no copying of memory. 

libRoadRunner extends the NumPy array to contain row and column name information and to support access to rows and columns by name, and extends the NumPy print method to format the extended arrays for printing. Since libRoadRunner users  frequently need to display the components and interaction names in the stoichiometry matrix, the libRoadRunner API makes this display is a single line task:
\begin{lstlisting}[language=Python]
print(r.getFullStoichiometryMatrix()) 
     J0, J1, J2, J3, J4 
S1 [[ 1, -1,  0,  0,  0], 
S2  [ 0,  1, -1,  0,  0], 
S3  [ 0,  0,  1, -1,  0], 
S4  [ 0,  0,  0,  1, -1]].
\end{lstlisting}
Running a simulation is straight forward and only requires users to load a model and call the simulation method. Defaults are in place that include time spans and number of points generated during a simulation. By default the simulate method will return time in the first column and all floating model species in the remaining columns:
\begin{lstlisting}[language=Python]
r = RoadRunner("glycolysis.sbml") 
m = r.simulate(plot=True)
\end{lstlisting}
Here \texttt{m} is a NumPy array, and the optional \texttt{plot=True} argument to the \texttt{simulate} method calls the standard plotting library, matplotlib, to display a basic time-series plot of the simulation result.  We can customize the simulation parameters using optional arguments, \textit{e.g.}, to generate a $100$ data-point time series for parameter ``P1'' and concentration ``S1'' from an SBML-specified model between time $t=0$ and time $t=12$, we specify:
\begin{lstlisting}[language=Python]
r = RoadRunner("glycolysis.sbml") 
m = r.simulate(0, 10, 100, ['time', 'P', '[S1]'])
\end{lstlisting}
A variety of other built-in symbols are provided to access information on reaction rates, rates of change, eigenvalues etc. Like a MATLAB top-level function, the libRoadRunner \texttt{simulate} method provides a consistent front end to all libRoadRunner's integration engines. Since MATLAB is familiar to many scientists, a MATLAB-like architecture should reduce the effort to learn the libRoadRunner API. To simplify generation of simulation documentation, libRoadRunner methods support internal pydoc strings which interactive Python environments such as IPython or Tellurium~\citep{url:tellurium}, make available as pop-up hints.

In analogy with MATLAB, the libRoadRunner API uses dynamic Python object properties. Loading an SBML-specified model via libRoadRunner automatically adds the SBML model's symbol names to the \texttt{RoadRunner} object, allowing dynamic introspection and modification of the object. If a model contains parameters and species \texttt{{'x', 'y', 'S1', 'S2'}}, the RoadRunner object will include these names as properties, which a user can read or set. \textit{E.g.}, 
\begin{lstlisting}[language=Python]
# load a model that has ids 'x', 'y' and 'S1'
r = RoadRunner('some_model.xml') 
r.x = 1.5    # set the 'x' parameter to 1.5         
r.y = 2.0    # set the 'y' parameter to 2.0
print(r.S1)  # print the 'S1' species concentration
\end{lstlisting}

\subsection*{Support for Analysis}
Since libRoadRunner was inspired and developed from the original C\# roadrunner, libRoadRunner inherits many of the analysis functions that roadRunner has. This includes methods to calculate both scaled and unscaled control coefficients, elasticities, rates to changes in all parameters, including conserved quantities, eigenvalues and eigenvectors and a variety of stoichiometric results such as the Link and K matrices~\citep{reder1988}. libRoadRunner can also compute the frequency response in order to make Bode plots. 

\subsection*{Identification of Conserved Quantities}
Many  computations of biochemical networks requires identification of conserved quantities (moieties in biochemical usage) and elimination of linearly-dependent species to avoid inversion of singular Jacobian matrices~\citep{Vallabhajosyula:2006vr}. libRoadRunner implements a libSBML plug-in which performs this reduction on a SBML Document object, first identifying conserved quantities and dependent species, then adding the conserved quantities to the document as set of global parameters and replacing the dependent species with assignment rules. The user can modify these conserved quantities, which behave as parameters, to investigate their effect on the dynamics of the model.

\subsection*{Hybrid Models}

libRoadRunner includes three independent simulation approaches, deterministic, stochastic and Flux Balance Analysis~\citep{orth2010flux}. As a result libRoadRunner is ideally suited for simulating hybrid models such as the whole-cell hybrid model by Karr~\citep{karr2012whole} where the different methods can be combined through python connecting code. 

\section{Use Cases}
libRoadRunner's ease of use, ability to handle very complex SBML models and very fast model execution speed have led to its rapid adoption in a variety of applications. 

\subsection*{Interactive Application}
A cross-platform integrated Python environment called Tell\-urium~\citep{url:tellurium} has been developed based on the Spyder IDE~\citep{Spyder2IDE:2014}, which combines libRoadRunner, libSBML, Antimony~\citep{smith2009antimony}, libSEDML and a variety of other packages. Tellurium provides a comprehensive development and analysis environment for SBML-specified models. Tellurium uses Antimony as a model specification language and libRoadRunner provides simulation execution and analysis functions. While libRoadRunner's execution speed is not a priority in this application, the concise syntax and intuitiveness of its Python API are essential to Tellurium's convenient interactive creation, simulation and analysis of DMN models. A detailed description of Tellurium will be given in a separate publication.  

\subsection*{Integrating SBML-model specifications into Multicell Models simulated in CompuCell3D}
CompuCell3D (\textit{CC3D}), a simulation environment for multiscale, multicell virtual-tissue model development and simulation, was the first tool to adopt libRoadRunner as a core engine. CC3D proper defines an object class of cell agent and behavior methods to allow cell agents to grow, divide, die, secrete/absorb chemicals, move, \textit{etc.}... The libRoadRunner/CC3D integration allows the state of an SBML-specified model inside a cell object to control the CC3D parameters describing the cell agent's behaviors and \textit{vice versa}. CC3D also uses libRoadRunner to time-step the state of a biochemical models and handles their interactions with other model objects via Python. 

\textit{E.g.}, in a model of changes in cell-cell adhesion leading to invasive tumor phenotypes, the CC3D cell objects have a CC3D parameter \textit{adhesion-molecule density}, which controls the CC3D behavior \textit{cell-cell adhesion}. An SBML-specified model relates the level of the transmembrane adhesion receptor E-cadherin in each cell to the cells' level of $\beta$-catenin~\citep{Andasari:2012wo}. The CC3D-model specification uses the libRoadRunner Python API to connect the CC3D \textit{adhesion-molecule density} to the SBML-model transmembrane E-Cadherin level. At run-time, libRoadRunner engine time evolves the model inside the cells, while the CC3D engine handles the evolution of the cell objects. 

Another example of the use of SBML models in virtual-tissue modeling is simulation of Delta-Notch patterning during embryonic development. Delta and Notch are heterophilic transmembrane receptors whose signaling is mutually inhibitory within a cell. However, the level of signaling depends on both the amount of Delta on the membrane of a cell and the amount of Notch on the surfaces of neighboring cells and \textit{vice versa}. Thus, to understand the dynamics of the signaling network, we need to know not only the model within the cell, but the pattern of its contacts with neighboring cells and their levels of Delta and Notch. To model this situation, we create CC3D cell objects and arrange them in an epithelium (a quasi-2D sheet). Each cell contains an SBML-specified model that describes how the cell's levels of membrane-bound and cytosolic Delta and Notch change, given a particular input level of transmembrane Delta and Notch signaling ~\citep{Swat:2012iza}. A Python layer uses the libRoadRunner API to combine the amount of Delta on the membrane of each cell, the amount of Notch on the membrane of each adjacent cell (adjacency is a CC3D model parameter) and the (CC3D model) contact area between each adjoining pair of cells to calculate the strength of Delta and Notch signaling each cell experiences. libRoadRunner then updates the internal state of the Delta-Notch signaling and regulatory networks using these signaling strengths as boundary conditions, while CC3D updates the cellshapes and positions and identifies cell rearrangements changing adjacencies and contact areas between cells. Together, the interactions produce the typical checkerboard pattern of embryonic Delta-Notch signaling.

\subsection*{Multi-scale Modeling of Liver Metabolism using libRoadRunner for Numerical Integration}
Liver function arises as a complex interaction among morphology, perfusion, and metabolism on multiple scales. \textit{E.g.}, the Virtual Liver Network has developed an organ-level model of human galactose clearance which includes single cell metabolism of hepatocytes, a representation of ultra-structure and micro-circulation in hepatic tissue, and a description of the structure of the entire organ to investigate how galactose clearance on organ level results from the interplay of blood flow, tissue structure and cellular metabolism~\citep{url:konigliver}.
  
The liver model includes an SBML-specified model of the smallest functional unit of the liver, a sinusoid consisting of a perfused capillary surrounded by hepatocytes. The sinusoid model contains a biochemical network describing galactose metabolism in individual hepatocytes, coupled via SBML-specified discretized transport equations for convection and diffusion, resulting in a model with several thousand components and interactions. The model also uses SBML events to describe the time-varying supply of galactose to the liver. To account for heterogeneity in blood flow and tissue architecture required simulation of more than $10^5$ (around 100,000 - 200,000 simulations necessary in total) replicas of the model under varying tissue and flow parameters, which became feasible only due to libRoadRunner's fast time-series generation and support for variable step sizes, which dramatically reduced output file size. libRoadRunner's Python API supported rewrite-free integration of the SBML models into a complex pre-existing modeling workflow, which included data management using Django, model annotations using Python bindings to libSBML, model prototyping using Python bindings to antimony and visualization of results using the Python REST interface to Cytoscape (CySBML, CyFluxViz).

\subsection*{Modeling of Synaptic, Neuronal and Neuron Network Dynamics in MEMORY platform using libRoadRunner} 
The MEMORY platform (Multiscale intEgrated Model Of the neRvous sYstem, formerly EONS~\citep{Bouteiller:2008kk}) simulates the function and dynamics of elements ranging from single channels or receptors (\textit{elementary models}), to synapses, which include many elementary models, to neurons, which themselves may include a large number of synapses. \textit{E.g.}, an SBML-specified neuron model may include many SBML-specified synapse models, each of which includes many SBML-specified neurotransmitter release and diffusion, AMPA receptor and NMDA-receptor models (both ionotropic receptors for the glutamate neurotransmitter). MEMORY depends on the flexibility and ease of use of libRoadRunner to enable the assembly of such complex hierarchical models. Neuronal models may be large, \textit{e.g.}, representing 10 ionotropic synapses in a CA1 neuron model~\citep{Izhikevich:2003wq} requires 73 events, 290 reactions, 414 rules and 1459 parameters. The speed of libRoadRunner time-series generation enables MEMORY to solve complex neuronal models quickly. 

Additionally, to ensure that a neuronal model quantitatively predicts biological functions like membrane potentials or intracellular molecular concentrations, MEMORY can optimize the model's parameters by fitting between multiple simulation and experimental time-series for characteristics including changes in receptor conductance, desensitization properties and spiking patterns. MEMORY uses evolutionary multi-objective optimization (from the EMOO framework~\citep{Bahl:2012uh}), which requires large numbers of simulation replicas. \textit{E.g.}, elementary-model optimization of an NMDA-receptor model with respect to eight distinct experimental results for dynamical changes in receptor channel conductance following paired-pulse stimulation, required 15,000 generations with 400 individuals per generation, \textit{i.e.}, 6 million simulation replicas (corresponding to 13,000 hours of simulated time). libRoadRunner took 66 hours to run the entire optimization on a 400-node computer cluster, orders of magnitude faster than other SBML simulators~\citep{bouteiller:2015}. 
\section{Conclusions} 
\label{ch:conclusion}
libRoadRunner's speed and ease of integration allow researchers to solve very large models, to include models in multiscale systems and run large ensembles of smaller models. The use of Python makes simulations easy to carry out for new users, while C++ and C APIs are attractive to developers looking to integrate libRoadRunner's capabilities into their existing simulation frameworks. libRoadRunner runs on both x86 and ARM architectures. We have successfully deployed libRoadrunner to Windows, Mac OS X, Linux, Raspberry Pi, NVIDIA Jetson TK1 and Parallella boards. 
\section{Future Work}
Our roadmap currently includes the following proposed extensions to libRoadRunner.

\paragraph{Support Automatic Differentiation of Jacobian calculations} 
lib\-Road\-Runner's stability and time evolution modules require numerical differentiation of the evolution equations to calculate a Jacobian matrix. For stiff equations, numerical differentiation requires the use of slow implicit solvers, which can introduce round-off errors. While we could sometimes calculate the Jacobian analytically from the AST, symbolic differentiation can produce equations with impractically many terms~\citep{Sauro:1993fm}. We are therefore evaluating an intermediate solution, known as automatic differentiation~(\textit{AD})~\citep{griewank1989automatic}, which computes analytic derivatives of programs rather than mathematical expressions. Simple analytical derivatives exist for the elementary mathematical functions (such as $\sin$, $\cos$, $\exp$, \textit{etc.}) available in most programming languages. Since most functions a program computes are compositions of these intrinsic functions, their derivatives are chain-combinations of the elementary derivatives. A JIT compilation of the resulting program for the Jacobian is usually much more compact than the equivalent analytical expansion.

\paragraph{Improve Steady-State Solvers} 
libRoadRunner uses the FORT\-RAN NLEQ2 nonlinear steady-state solver, which is not \textit{thread safe} (multiple  execution threads may call a routine concurrently). The remainder of the libRoadRunner library is thread-safe, so we had to place exclusive access locks (mutexes) on the NLEQ solver which restricts its use to one thread at a time. To eliminate this restriction, we plan to add several thread-safe steady state solvers, including Sundials' KINSOL~\citep{hindmarsh2005sundials}, and a damped Newton algorithm~\citep{Hoops:2006ui, Sauro:1993fm}.

\paragraph{Add Additional Integrators} 
libRoadRunner includes the explicit and implicit integrators from the Sundials suite, a Runge-Kutta integrator and a Gillespie integrator. We plan to add  integrators based on LSODA, ARKode~\citep{arkode:2014} for the Sundials suite, and the banded integrator from CVODE (plus a tool to detect Jacobians with banded structure). The CVODE integrator is much faster and uses much less memory than other integrators when solving models with banded Jacobians.

\paragraph{JDesigner} A new crossplaform version of JDesigner is under development and will employ libRoadRunner as its simulation engine~\citep{url:pathwaydesigner}. 

\section{Acknowledgments} 
Endre T. Somogyi, Maciej H. Swat and James A. Glazier acknowledge support from NIH grants R01
GM077138, U01 GM111243 and R01 GM076692 and EPA grant RD83500101.  Matthias K\"{o}nig acknowledges
support from the Federal Ministry of Education and Research (BMBF, Germany) within the Virtual Liver
Network (VLN grant 0315741). Jean-Marie Bouteiller acknowledges support from NIH grants P41 EB001978
and U01 GM104604. Herbert M. Sauro acknowledges support from NIH grant R01 GM081070. The content is
solely the responsibility of the authors and does not necessarily represent the official views of
the National Institutes of Health. We would also  like to acknowledge Totte Karlsson for the
original C\# to C++ translation, C compiler backend and C API, Stanley Gu for testing the library as
a web service, Lucian Smith for developing a major part of the test suite, Mike Galdzicki for
developing the detailed build instructions for developers. HS conceived the project, helped with
documentation, design and testing, AS designed the overall architecture and JIT compiler, MS for Linux testing and builds, MK and JB for carrying out extensive testing, KM for the ARM based development and JG for editing the manuscript.

\bibliographystyle{natbib}

\bibliography{papers,bibdesk}

\end{document}

%% file: preamble.tex
\usepackage{setspace}
\usepackage{extarrows}
\usepackage{algorithm}
\usepackage{algorithmic}
\usepackage{longtable}
\usepackage{tikz}
\usepackage{amsmath, amsthm, amssymb}
\usepackage{tabularx}
\usepackage[version=3]{mhchem}
\usepackage{url}
\usepackage[procnames]{listings}
\usepackage{todonotes}

%% file: Main.bbl
\begin{thebibliography}{}

\bibitem[Ackermann {\em et~al.}(2009)Ackermann, Baecher, Franzel, Goesele, and
  al]{Ackermann:2009vq}
Ackermann, J., Baecher, P., Franzel, T., Goesele, M., and al, e. (2009).
\newblock {Massively-Parallel Simulation of Biochemical Systems.}
\newblock {\em GI Jahrestagung\/}.

\bibitem[Alfred V.~Aho(1986)Alfred V.~Aho]{Aho:1986}
Alfred V.~Aho, Ravi~Sethi, J. D.~U. (1986).
\newblock {\em Compilers: principles, techniques, and tools\/}.
\newblock Addison-Wesley Longman Publishing Co.

\bibitem[Andasari {\em et~al.}(2012)Andasari, Roper, Swat, and
  Chaplain]{Andasari:2012wo}
Andasari, V., Roper, R.~T., Swat, M.~H., and Chaplain, M.~A. (2012).
\newblock {Integrating intracellular dynamics using CompuCell3D and
  Bionetsolver: applications to multiscale modelling of cancer cell growth and
  invasion}.
\newblock {\em PLoS ONE\/}, {\bf 7}(3), e33726.

\bibitem[Bahl {\em et~al.}(2012)Bahl, Stemmler, Herz, and Roth]{Bahl:2012uh}
Bahl, A., Stemmler, M.~B., Herz, A.~V., and Roth, A. (2012).
\newblock {Automated optimization of a reduced layer 5 pyramidal cell model
  based on experimental data}.
\newblock {\em Journal of Neuroscience Methods\/}, {\bf 210}(1), 22--34.

\bibitem[Beazley {\em et~al.}(1996)Beazley {\em et~al.}]{beazley1996swig}
Beazley, D.~M. {\em et~al.} (1996).
\newblock {SWIG: An easy to use tool for integrating scripting languages with C
  and C++}.
\newblock In {\em {Proceedings of the 4th USENIX Tcl/Tk workshop}\/}, pages
  129--139.

\bibitem[Bergmann and Sauro(2006)Bergmann and Sauro]{Bergmann:2006ut}
Bergmann, F.~T. and Sauro, H.~M. (2006).
\newblock {SBW - A modular framework for systems biology}.
\newblock In {\em WSC '06 Proceedings of the 38th conference on Winter
  simulation\/}, pages 1637--1645. Winter Simulation Conference.

\bibitem[{BioModels.net Team}(2014){BioModels.net Team}]{url:biomodels}
{BioModels.net Team} (2014 (accessed Nov 14, 2014)).
\newblock {BioModels Database}.
\newblock \url{http://www.ebi.ac.uk/biomodels-main/}.

\bibitem[Bornstein {\em et~al.}(2008)Bornstein, Keating, Jouraku, and
  Hucka]{bornstein2008libsbml}
Bornstein, B.~J., Keating, S.~M., Jouraku, A., and Hucka, M. (2008).
\newblock {LibSBML: an API library for SBML}.
\newblock {\em Bioinformatics\/}, {\bf 24}(6), 880--881.

\bibitem[Bouteiller {\em et~al.}(2014)Bouteiller, Feng, Onopa, Huang, Hu,
  Somogyi, Baudry, Bischoff, and Berger]{bouteiller:2014}
Bouteiller, J., Feng, Z., Onopa, A., Huang, E., Hu, E., Somogyi, E., Baudry,
  M., Bischoff, S., and Berger, T. (2014).
\newblock Maximizing predictability of a bottom-up complex multi-scale model
  through systematic validation and multi-objective multi-level optimization.
\newblock In {\em Neuroscience 2014\/}, Washington D.C. Society for
  Neuroscience.

\bibitem[Bouteiller {\em et~al.}(2015)Bouteiller, Feng, Onopam, Huang, Hu,
  Somogyi, Baudry, Bischoff, and Berger]{bouteiller:2015}
Bouteiller, J.-M., Feng, Z., Onopam, A., Huang, M., Hu, E.~Y., Somogyi, E.,
  Baudry, M., Bischoff, S., and Berger, T.~W. (2015).
\newblock {Maximizing Predictability of a Bottom-Up Complex Multi-Scale Model
  through Systematic Validation and Multi-Objective Multi-Level Optimization}.
\newblock In {\em Proceedings of the 7th International IEEE/EMBS Conference on
  Neural Engineering (NER)\/}.

\bibitem[Bouteiller {\em et~al.}(2008)Bouteiller, Baudry, Allam, GREGET,
  Bischoff, and Berger]{Bouteiller:2008kk}
Bouteiller, J.-M.~C., Baudry, M., Allam, S.~L., GREGET, R.~J., Bischoff, S.,
  and Berger, T.~W. (2008).
\newblock {MODELING GLUTAMATERGIC SYNAPSES: INSIGHTS INTO MECHANISMS REGULATING
  SYNAPTIC EFFICACY}.
\newblock {\em Journal of Integrative Neuroscience\/}, {\bf 07}(02), 185--197.

\bibitem[Griewank {\em et~al.}(1989)Griewank {\em
  et~al.}]{griewank1989automatic}
Griewank, A. {\em et~al.} (1989).
\newblock On automatic differentiation.
\newblock {\em Mathematical Programming: recent developments and
  applications\/}, {\bf 6}, 83--107.

\bibitem[Hester {\em et~al.}(2011)Hester, Belmonte, Gens, Clendenon, and
  Glazier]{Hester:2011tz}
Hester, S.~D., Belmonte, J.~M., Gens, J.~S., Clendenon, S.~G., and Glazier,
  J.~A. (2011).
\newblock {A multi-cell, multi-scale model of vertebrate segmentation and
  somite formation}.
\newblock {\em PLoS Computational Biology\/}, {\bf 7}(10), e1002155.

\bibitem[Hindmarsh {\em et~al.}(2005)Hindmarsh, Brown, Grant, Lee, Serban,
  Shumaker, and Woodward]{hindmarsh2005sundials}
Hindmarsh, A.~C., Brown, P.~N., Grant, K.~E., Lee, S.~L., Serban, R., Shumaker,
  D.~E., and Woodward, C.~S. (2005).
\newblock {SUNDIALS: Suite of nonlinear and differential/algebraic equation
  solvers}.
\newblock {\em {ACM Transactions on Mathematical Software (TOMS)}\/}, {\bf
  31}(3), 363--396.

\bibitem[Hofmeyr and Cornish-Bowden(1997)Hofmeyr and
  Cornish-Bowden]{ReversibleHill:1997}
Hofmeyr, J.-H.~S. and Cornish-Bowden, A. (1997).
\newblock The reversible hill equation: how to incorporate cooperative enzymes
  into metabolic models.
\newblock {\em Comput Appl Biosci\/}, {\bf 13(4)}, 377--385.

\bibitem[Hoops {\em et~al.}(2006)Hoops, Sahle, Gauges, Lee, Pahle, Simus,
  Singhal, Xu, Mendes, and Kummer]{Hoops:2006ui}
Hoops, S., Sahle, S., Gauges, R., Lee, C., Pahle, J., Simus, N., Singhal, M.,
  Xu, L., Mendes, P., and Kummer, U. (2006).
\newblock {COPASI---a complex pathway simulator}.
\newblock {\em Bioinformatics\/}, {\bf 22}(24), 3067--3074.

\bibitem[Hucka {\em et~al.}(2003)Hucka, Finney, Sauro, Bolouri, Doyle, Kitano,
  , the rest of~the SBML~Forum, Arkin, Bornstein, Bray, Cornish-Bowden,
  Cuellar, Dronov, Gilles, Ginkel, Gor, Goryanin, Hedley, Hodgman, Hofmeyr,
  Hunter, Juty, Kasberger, Kremling, Kummer, Le~Novere, Loew, Lucio, Mendes,
  Minch, Mjolsness, Nakayama, Nelson, Nielsen, Sakurada, Schaff, Shapiro,
  Shimizu, Spence, Stelling, Takahashi, Tomita, Wagner, and Wang]{Hucka:2003fs}
Hucka, M., Finney, A., Sauro, H.~M., Bolouri, H., Doyle, J.~C., Kitano, H., ,
  the rest of~the SBML~Forum, Arkin, A.~P., Bornstein, B.~J., Bray, D.,
  Cornish-Bowden, A., Cuellar, A.~A., Dronov, S., Gilles, E.~D., Ginkel, M.,
  Gor, V., Goryanin, I.~I., Hedley, W.~J., Hodgman, T.~C., Hofmeyr, J.~H.,
  Hunter, P.~J., Juty, N.~S., Kasberger, J.~L., Kremling, A., Kummer, U.,
  Le~Novere, N., Loew, L.~M., Lucio, D., Mendes, P., Minch, E., Mjolsness,
  E.~D., Nakayama, Y., Nelson, M.~R., Nielsen, P.~F., Sakurada, T., Schaff,
  J.~C., Shapiro, B.~E., Shimizu, T.~S., Spence, H.~D., Stelling, J.,
  Takahashi, K., Tomita, M., Wagner, J., and Wang, J. (2003).
\newblock {The systems biology markup language (SBML): a medium for
  representation and exchange of biochemical network models}.
\newblock {\em Bioinformatics\/}, {\bf 19}(4), 524--531.

\bibitem[Izhikevich(2003)Izhikevich]{Izhikevich:2003wq}
Izhikevich, E.~M. (2003).
\newblock {Simple model of spiking neurons}.
\newblock {\em Neural Networks, IEEE Transactions on\/}, {\bf 14}(6),
  1569--1572.

\bibitem[Karr {\em et~al.}(2012)Karr, Sanghvi, Macklin, Gutschow, Jacobs,
  Bolival, Assad-Garcia, Glass, and Covert]{karr2012whole}
Karr, J.~R., Sanghvi, J.~C., Macklin, D.~N., Gutschow, M.~V., Jacobs, J.~M.,
  Bolival, B., Assad-Garcia, N., Glass, J.~I., and Covert, M.~W. (2012).
\newblock A whole-cell computational model predicts phenotype from genotype.
\newblock {\em Cell\/}, {\bf 150}(2), 389--401.

\bibitem[Keller {\em et~al.}(2013)Keller, D{\"o}rr, Tabira, Funahashi, Ziller,
  Adams, Rodriguez, Nov{\`e}re, Hiroi, Planatscher, Zell, and
  Dr{\"a}ger]{Keller:2013cn}
Keller, R., D{\"o}rr, A., Tabira, A., Funahashi, A., Ziller, M.~J., Adams, R.,
  Rodriguez, N., Nov{\`e}re, N.~L., Hiroi, N., Planatscher, H., Zell, A., and
  Dr{\"a}ger, A. (2013).
\newblock {The systems biology simulation core algorithm}.
\newblock {\em BMC Systems Biology\/}, {\bf 7}(1), 55.

\bibitem[K\"{o}nig(2015)K\"{o}nig]{url:konigliver}
K\"{o}nig, M. (2015 (accessed Feb 8, 2015)).
\newblock {Multiscale model of hepatic galactose metabolism}.
\newblock \url{https://github.com/matthiaskoenig/multiscale-galactose}.

\bibitem[Lattner and Adve(2004)Lattner and Adve]{Lattner:2004vw}
Lattner, C. and Adve, V. (2004).
\newblock {LLVM: A compilation framework for lifelong program analysis {\&}
  transformation}.
\newblock In {\em Code Generation and Optimization 2004\/}, pages 75--86. IEEE.

\bibitem[Moraru {\em et~al.}(2008)Moraru, Morgan, Li, Loew, Schaff,
  Lakshminarayana, Slepchenko, Gao, and Blinov]{Moraru:2008iv}
Moraru, I.~I., Morgan, F., Li, Y., Loew, L.~M., Schaff, J.~C., Lakshminarayana,
  A., Slepchenko, B.~M., Gao, F., and Blinov, M.~L. (2008).
\newblock {Virtual Cell modelling and simulation software environment}.
\newblock {\em IET Systems Biology\/}, {\bf 2}(5), 352--362.

\bibitem[Myers {\em et~al.}(2009)Myers, Barker, Jones, Kuwahara, Madsen, and
  Nguyen]{iBioSim:2009}
Myers, C.~J., Barker, N., Jones, K., Kuwahara, H., Madsen, C., and Nguyen,
  N.~P. (2009).
\newblock {i{B}io{S}im: a tool for the analysis and design of genetic
  circuits}.
\newblock {\em Bioinformatics\/}, {\bf 25}(21), 2848--2849.

\bibitem[Nowak and Weimann(1991)Nowak and Weimann]{nowak1991family}
Nowak, U. and Weimann, L. (1991).
\newblock A family of newton codes for systems of highly nonlinear equations.
\newblock Technical report, Citeseer.

\bibitem[Olivier and Bergmann(2012)Olivier and Bergmann]{url:sbmlfbc}
Olivier, B. and Bergmann, F. (2012).
\newblock {Flux Balance Constraints (fbc)}.
\newblock
  \url{http://sbml.org/Community/Wiki/SBML_Level_3_Proposals/Flux_Constraints}.
\newblock (accessed February 23, 2015).

\bibitem[Olivier {\em et~al.}(2005)Olivier, Rohwer, and Hofmeyr]{Pysces:2005}
Olivier, B.~G., Rohwer, J.~M., and Hofmeyr, J.~H. (2005).
\newblock {{M}odelling cellular systems with {P}y{S}{C}e{S}}.
\newblock {\em Bioinformatics\/}, {\bf 21}(4), 560--561.

\bibitem[Orth {\em et~al.}(2010)Orth, Thiele, and Palsson]{orth2010flux}
Orth, J.~D., Thiele, I., and Palsson, B.~{\O}. (2010).
\newblock What is flux balance analysis?
\newblock {\em Nature biotechnology\/}, {\bf 28}(3), 245--248.

\bibitem[Raybaut and Cordoba(2015)Raybaut and Cordoba]{Spyder2IDE:2014}
Raybaut, P. and Cordoba, C. (2015 (accessed February 3, 2015)).
\newblock {\em Spyder is the Scientific PYthon Development EnviRonment\/}.
\newblock \url{http://code.google.com/p/spyderlib/}.

\bibitem[Reder(1988)Reder]{reder1988}
Reder, C. (1988).
\newblock Metabolic control theory: a structural approach.
\newblock {\em Journal of Theoretical Biology\/}, {\bf 135}(2), 175--201.

\bibitem[Reynolds {\em et~al.}(2014)Reynolds, Woodward, Gardner, and
  Hindmarsh]{arkode:2014}
Reynolds, D.~R., Woodward, C.~S., Gardner, D.~J., and Hindmarsh, A.~C. (2014).
\newblock {ARKode: A library of high order implicit/explicit methods for
  multi-rate problems}.
\newblock In {\em {SIAM Conference on Parallel Processing for Scientific
  Computing}\/}.

\bibitem[Romer {\em et~al.}(1996)Romer, Lee, Voelker, Wolman, Wong, Baer,
  Bershad, and Levy]{Romer:1996wj}
Romer, T.~H., Lee, D., Voelker, G.~M., Wolman, A., Wong, W.~A., Baer, J.-L.,
  Bershad, B.~N., and Levy, H.~M. (1996).
\newblock {The structure and performance of interpreters}.
\newblock {\em ACM SIGPLAN Notices\/}, {\bf 31}(9), 150--159.

\bibitem[Sauro(1993)Sauro]{Sauro:1993fm}
Sauro, H.~M. (1993).
\newblock {SCAMP: a general-purpose simulator and metabolic control analysis
  program}.
\newblock {\em Bioinformatics\/}, {\bf 9}(4), 441--450.

\bibitem[Sauro(2012)Sauro]{SauroBookEnzymeKinetics}
Sauro, H.~M. (2012).
\newblock {\em Enzyme Kinetics for Systems Biology\/}.
\newblock Ambrosius Publishing.

\bibitem[Sauro(2014)Sauro]{SauroBookPathwayModeling}
Sauro, H.~M. (2014).
\newblock {\em Systems Biology: An Introduction to Pathway Modeling\/}.
\newblock Ambrosius Publishing.

\bibitem[Sauro(2015)Sauro]{url:pathwaydesigner}
Sauro, H.~M. (2014 (accessed Feb 12, 2015)).
\newblock {pathwayDesginer: Visual Modeling Environment for cellular Systems}.
\newblock http://pathwaydesigner.org/.

\bibitem[Sauro and Fell(2000)Sauro and Fell]{sauro2000jarnac}
Sauro, H.~M. and Fell, D.~A. (2000).
\newblock Jarnac: a system for interactive metabolic analysis.
\newblock In {\em Animating the Cellular Map: Proceedings of the 9th
  International Meeting on BioThermoKinetics\/}, pages 221--228. Stellenbosch
  University Press.

\bibitem[Sauro {\em et~al.}(2014)Sauro, Galdzicki, Somogyi, Karlsson, Smith,
  Gu, Darling, Elmi, and Stocking]{url:tellurium}
Sauro, H.~M., Galdzicki, M., Somogyi, A., Karlsson, T., Smith, L., Gu, S.,
  Darling, A., Elmi, N., and Stocking, K. (2014 (accessed March 5, 2014)).
\newblock {Model, simulate, and analyze biochemical systems using a single
  tool}.
\newblock \url{http://tellurium.analogmachine.org/}.

\bibitem[Smith {\em et~al.}(2009)Smith, Bergmann, Chandran, and
  Sauro]{smith2009antimony}
Smith, L.~P., Bergmann, F.~T., Chandran, D., and Sauro, H.~M. (2009).
\newblock Antimony: a modular model definition language.
\newblock {\em Bioinformatics\/}, {\bf 25}(18), 2452--2454.

\bibitem[{SOSLib Team}(2014){SOSLib Team}]{url:soslib}
{SOSLib Team} (2014 (accessed March 5, 2014)).
\newblock {The SBML ODE Solver Library}.
\newblock \url{http://www.tbi.univie.ac.at/~raim/odeSolver/news/}.

\bibitem[Swat {\em et~al.}(2012)Swat, Thomas, Belmonte, Shirinifard, Hmeljak,
  and Glazier]{Swat:2012iza}
Swat, M.~H., Thomas, G.~L., Belmonte, J.~M., Shirinifard, A., Hmeljak, D., and
  Glazier, J.~A. (2012).
\newblock {Multi-Scale Modeling of Tissues Using CompuCell3D}.
\newblock In {\em Methods in cell biology\/}, pages 325--366. Elsevier.

\bibitem[Takizawa {\em et~al.}(2013)Takizawa, Nakamura, Tabira, Chikahara,
  Matsui, Hiroi, and Funahashi]{Takizawa:2013tj}
Takizawa, H., Nakamura, K., Tabira, A., Chikahara, Y., Matsui, T., Hiroi, N.,
  and Funahashi, A. (2013).
\newblock {LibSBMLSim: a reference implementation of fully functional SBML
  simulator}.
\newblock {\em Bioinformatics\/}, {\bf 29}(11), 1474--1476.

\bibitem[Vallabhajosyula {\em et~al.}(2006)Vallabhajosyula, Chickarmane, and
  Sauro]{Vallabhajosyula:2006vr}
Vallabhajosyula, R.~R., Chickarmane, V., and Sauro, H.~M. (2006).
\newblock {Conservation analysis of large biochemical networks}.
\newblock {\em Bioinformatics\/}, {\bf 22}(3), 346--353.

\end{thebibliography}
